\documentclass[seceq]{ptptex}

\usepackage{graphicx}

\makeatletter
\def\simleq{\mathrel{\mathpalette\gl@align<}}
\def\simgeq{\mathrel{\mathpalette\gl@align>}}
\def\gl@align#1#2{\lower.6ex\vbox{\baselineskip\z@skip\lineskip\z@
     \ialign{$\m@th#1\hfill##\hfil$\crcr#2\crcr\sim\crcr}}}
\makeatother

\newcommand{\bra}{\langle}
\newcommand{\ket}{\rangle}
\newcommand{\braket}[1]{\bra #1 \ket}
\newcommand{\qq}{\braket{\bar{q}q}}

\newcommand{\qGq}{g\braket{\bar{q}\sigma_{\mu\nu}G_{\mu\nu} q}}

\title{
The Determination of 
the Quark-Gluon Mixed Condensate 
$g\langle\bar{q}\sigma_{\mu\nu}G_{\mu\nu} q\rangle$
from Lattice QCD
}

\author{
Takumi \textsc{Doi}$^{1,}$\footnote{E-mail: doi@th.phys.titech.ac.jp}, 
Noriyoshi \textsc{Ishii}$^2$, 
Makoto \textsc{Oka}$^1$ and 
Hideo \textsc{Suganuma}$^1$
}

\inst{
$^1$ Department of Physics, Tokyo Institute of Technology, Tokyo 152-8551, Japan\\
$^2$ Radiation Laboratory, The Institute of Physical and Chemical Research (RIKEN),
Wako, 351-0198, Japan
}

\abst{
We study the quark-gluon mixed condensate 
$g\langle\bar{q}\sigma_{\mu\nu}G_{\mu\nu} q\rangle$,
using the SU(3)$_c$ lattice QCD with the Kogut-Susskind fermion 
at the quenched level.
We generate 100 gauge configurations
on the $16^4$ lattice with $\beta = 6.0$, and perform the 
measurement of 
$g\langle\bar{q}\sigma_{\mu\nu}G_{\mu\nu} q\rangle$
at 16 points in each gauge configuration
for each current quark mass of $m_q=21, 36, 52$ MeV.
Using the 1600 data for each $m_q$, 
we find 
$m_0^2 \equiv 
g\langle\bar{q}\sigma_{\mu\nu}G_{\mu\nu} q\rangle / 
\langle\bar{q}q\rangle
\simeq 2.5$ GeV$^2$ 
at the lattice scale of
$a^{-1} \simeq 2 {\rm GeV}$ in the chiral limit.
The large value of 
$g\langle\bar{q}\sigma_{\mu\nu}G_{\mu\nu} q\rangle$
suggests its importance in
the operator product expansion in QCD.
We study also chiral restoration at finite temperature
in terms of 
$g\langle\bar{q}\sigma_{\mu\nu}G_{\mu\nu} q\rangle$,
which is another chiral order parameter.
We present the lattice QCD results of
$g\langle\bar{q}\sigma_{\mu\nu}G_{\mu\nu} q\rangle$
at finite temperature.
}

\begin{document}

\maketitle

\section{The importance of the quark-gluon mixed condensate $\qGq$}
\label{sec:intro}

The non-perturbative nature of 
quantum chromodynamics (QCD) 
is characterized by its nontrivial vacuum structure
such as various condensates.
%
%
Among various condensates, we emphasize here the importance 
of the quark-gluon mixed condensate 
$\qGq \equiv 
{g\braket{\bar{q}\sigma_{\mu\nu}G_{\mu\nu}^A \frac{1}{2}\lambda^A q}}$.
First, the mixed condensate represents a direct correlation 
between quarks and gluons in the QCD vacuum. 
In this point, the mixed condensate differs from $\qq$ and 
$\langle G_{\mu\nu}G^{\mu\nu} \rangle$ even at the qualitative level.
Second, this mixed condensate is another chiral order parameter 
of the second lowest dimension
and it flips the chirality of the quark as 
\begin{eqnarray}
\qGq = g\braket{\bar{q}_R\ (\sigma_{\mu\nu}G_{\mu\nu})\ q_L}
+ g\braket{\bar{q}_L\ (\sigma_{\mu\nu}G_{\mu\nu})\ q_R}.
\end{eqnarray}
Third, the mixed condensate plays an important role in
various QCD sum rules, especially in the baryons\cite{Ioffe,Dosch},
the light-heavy mesons~\cite{Dosch2}
and the exotic mesons~\cite{Latorre}.
In the QCD sum rules,
the value $m_0^2 \equiv \qGq / \qq \simeq 0.8\pm 0.2\ {\rm GeV}^2$ has been 
proposed as a result of the 
phenomenological analyses~\cite{Bel,RRY2,Ovchi,Narison1}.
However, in spite of the importance of $\qGq$, 
there was only one preliminary
lattice QCD study~\cite{K&S}, 
which was performed with insufficient
statistics (only 5 data)
using a small ($8^4$) 
and coarse lattice ($\beta=5.7$).

Therefore, we present the calculation of $\qGq$ in 
lattice QCD with a larger $(16^4)$ and finer $(\beta=6.0)$
lattice and with high statistics (1600 data).
We perform the measurement of $\qGq$ as well as $\qq$
in the SU(3)$_c$ lattice at the quenched level,
using the Kogut-Susskind (KS) fermion to respect chiral symmetry.
%
We generate 100 gauge configurations and pick up 16 space-time points 
for each configuration to calculate the condensates.
With this high statistics of 1600 data for each quark mass,
we perform a reliable estimate for 
the ratio $m_0^2\equiv \qGq/\qq$ at the lattice scale of 
$a^{-1} \simeq 2 {\rm GeV}$ 
in the chiral limit~\cite{DOIS:qGq}.

\section{Lattice Formalism}
\label{sec:formalism}

Since
both of the  condensates  $\qq$  and  $\qGq$
are chiral order parameters, 
they are sensitive to explicit chiral-symmetry breaking.
Therefore, we adopt the KS fermion, which
preserves the explicit chiral symmetry in massless quark limit, $m=0$.
On the other hand, the Wilson and  the clover fermions would not
appropriate for our study, because these fermions
explicitly break chiral symmetry even for $m=0$.

The action for the KS fermion is described by 
spinless Grassmann fields $\bar{\chi},\chi$
and the gauge link-variable $U_\mu \equiv \exp[ -iagA_\mu ]$.
When the gauge field is set to be zero,
the SU(4)$_f$ quark-spinor field $q$ with spinor $i$ and flavor $f$
is expressed 
by $\chi$ as 
\begin{eqnarray}
\label{eq:q-ks_trans}
q_i^f (x) 
&=&  \frac{1}{8}
\sum_{\rho}\
( \Gamma_\rho )_{if}\ 
\chi (x+\rho ), \ \ 
\Gamma_\rho \equiv \gamma_1^{\rho_1} \gamma_2^{\rho_2} \gamma_3^{\rho_3} \gamma_4^{\rho_4}, \ \ \rho \equiv (\rho_1,\rho_2,\rho_3,\rho_4)
\end{eqnarray}
where $\rho$ with $\rho_\mu \in \{0,1\}$ runs over the 16 sites 
in the $2^4$ hypercube.
When the gluon field is turned on, 
we insert
additional link-variables in Eq.~(\ref{eq:q-ks_trans})
in order to respect the gauge covariance.
Hence, the flavor-averaged condensates are expressed as 
\begin{eqnarray}
\label{eq:condensates-qq-def} 
&&a^3 \qq 
= - \frac{1}{4}\sum_f {\rm Tr}\left[ \braket{q^f(x) \bar{q}^f(x)} \right] 
%
%
= - \frac{1}{2^8}\sum_\rho
       {\rm Tr}\left[ \Gamma_\rho \Gamma_\rho^\dag\ 
	\braket{\chi(x+\rho ) \bar{\chi}(x+\rho )} \right], \\
\label{eq:condensates-qGq-def} 
&&a^5 \qGq
= - \frac{1}{4}\sum_f \sum_{\mu,\nu}{\rm Tr}
	\left[ \braket{q^f(x) \bar{q}^f(x)} \sigma_{\mu\nu} G_{\mu\nu}\right]  \nonumber \\
%
&&= - \frac{1}{2^8} \sum_{\mu,\nu} \sum_\rho
{\rm Tr}\left[\
  {\cal U}_{\pm\mu,\pm\nu}(x+\rho)\ 
\Gamma_{\rho'} \Gamma_{\rho}^\dag\
\braket{\chi (x+\rho') 
\bar{\chi}(x+\rho)} \
\sigma_{\mu\nu}\ 
  G_{\mu\nu}^{\rm lat}(x+\rho)\ 
\right], \nonumber \\ 
&& \\
&&  \qquad\qquad \rho' \equiv \rho \pm \mu \pm \nu, \nonumber
\end{eqnarray}
where the sign $\pm$ is taken such that the sink point 
$(x+\rho') = (x+\rho\pm\mu\pm\nu )$
belongs to the same hypercube of the source point $(x+\rho)$.
Here, in order to respect the gauge covariance,
we have used, in Eq.~(\ref{eq:condensates-qGq-def}),
\begin{eqnarray}
{\cal U}_{\pm\mu,\pm\nu}(x) \equiv 
        \frac{1}{2}\left[\ U_{\pm\mu} (x) U_{\pm\nu} (x\pm\mu ) 
                + U_{\pm\nu} (x) U_{\pm\mu} (x\pm\nu )\ \right],
\label{eq:qGq-def-U}
\end{eqnarray}
where the definition of $U_{-\mu} (x) \equiv U^\dag_{\mu}(x-\mu)$ is used.

%
%
On the gluon field strength $G_{\mu\nu}$,
we adopt the clover-type definition on the lattice,
\begin{eqnarray}
G_{\mu\nu}^{\rm lat}(s) = \frac{i}{16} \sum_A \lambda^A \ {\rm Tr}
\left[ 
 \lambda^A\{U_{\mu\nu}(s) +U_{\nu\,-\!\mu}(s)+U_{-\!\mu\,-\!\nu}(s)+U_{-\!\nu\,\mu}(s) \}
-\lambda^A \{\mu \leftrightarrow \nu \} 
\right], \nonumber \\
\label{eq:clover}
\end{eqnarray}
which contains no ${\cal O}(a)$ discretization error.
This is also an advanced point which is absent in Ref.~\citen{K&S}.


\section{The lattice QCD results}
\label{sec:results}

We calculate the condensates $\qq$ and $\qGq$ using the SU(3)$_c$ lattice QCD at the quenched level.
We perform the Monte Carlo simulation 
with the standard Wilson action at $\beta=6.0$ 
on the $16^4$ lattice.
%
%
The lattice unit $a\simeq 0.10\, {\rm fm}$ is obtained so as to reproduce 
the string tension $\sigma = 0.89\, {\rm GeV/fm}$~\cite{rabbit:3Q}.
We use the quark mass $m = 21, 36, 52$ MeV,
i.e., 
$ma = 0.0105,\ 0.0184,\ 0.0263$.
For the fields $\chi$, $\bar{\chi}$, the anti-periodic condition 
is imposed.
%
%
%
%
We measure the condensates 
on 16 different physical space-time points $x$ in each
configuration as $x=(x_1,x_2,x_3,x_4)$ with $x_\mu \in \{0, 8\}$ 
in the lattice unit.
%
%
For each $m$, we calculate the flavor-averaged condensates,
and average them over the 16 space-time points and 
100 gauge configurations.

\begin{figure}
\includegraphics[scale=0.6]{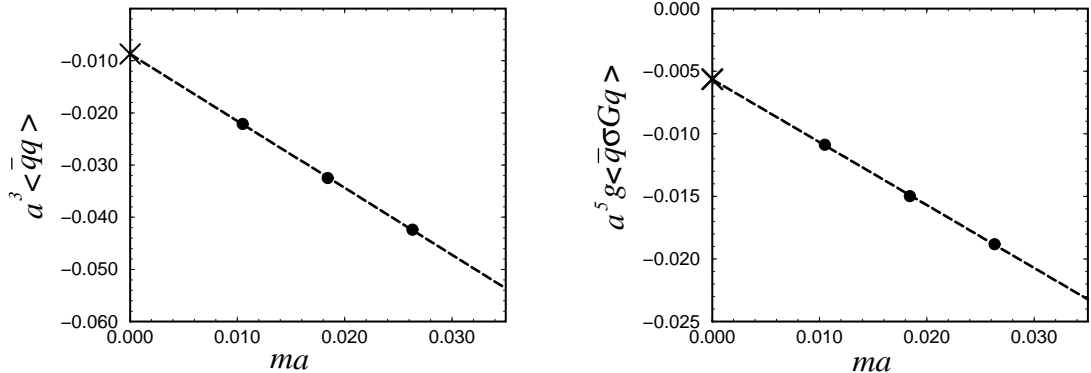}
\caption{\label{fig:plot}
The bare condensates $a^3\qq$ and $a^5\qGq$ plotted against the quark mass $ma$.
The dashed lines denote the best linear extrapolations, 
and the cross symbols correspond to the values in the chiral limit. 
The jackknife errors are hidden in the dots.
}
\end{figure}

Figure~\ref{fig:plot} shows the values of the
bare condensates $a^3\qq$ and $a^5\qGq$ against the quark mass $ma$.
We emphasize that the jackknife errors are almost negligible, 
due to the high statistics of $1600$ data for each quark mass.
From Fig.~\ref{fig:plot}, both $\qq$ and $\qGq$ show 
a clear linear behavior against the quark mass $m$.
Therefore, we fit the data with a linear function and determine 
the condensates in the chiral limit.
The results are summarized in Table~\ref{tab:mass-beta-6.0}.

\begin{table}[thb]
\caption{
The numerical results of $a^3\qq$ and $a^5\qGq$ for various $ma$.
The last column denotes their values in the chiral limit obtained 
by the linear chiral extrapolation.
\label{tab:mass-beta-6.0}
}
\begin{tabular}{ccccc}
\hline\hline
	   &  $ma=0.0263$     &  $ma=0.0184$     &  $ma=0.0105$      &  chiral limit\\
\hline
$a^3\qq$   &  $-0.04240(16)$  &  $-0.03247(15)$ &  $-0.02212(16)$  & 
$-0.00872(17)$ \\
$a^5\qGq$  &  $-0.01882(15)$  &  $-0.01498(14)$ &  $-0.01088(14)$  & 
$-0.00565(14)$ \\
\hline
\end{tabular}
\end{table}


We check the reliability of our lattice QCD results
by considering the finite volume artifact.
In order to estimate this artifact,
we carry out the same calculation imposing the periodic boundary condition 
on the Grassmann fields $\chi$ and $\bar{\chi}$,
instead of the anti-periodic boundary condition,
keeping the other parameters same.
%
We obtain that the results with different 
boundary conditions almost coincide within 
about 1\% difference, and thus we
conclude that the physical volume
$V \sim (1.6\ {\rm fm})^4$ in this simulation is large enough to 
avoid the finite volume artifact~\cite{DOIS:qGq}.
%
%



The values  of the  condensates in the  continuum limit  
are to be obtained after the renormalization,
which, however, suffers from the
uncertainty of the non-perturbative effect.
As a more reliable quantity,
we provide   the  ratio  $m_0^2  \equiv  \qGq   /  \qq$,
which is free  from the  uncertainty from  the wave
function  renormalization  of the  quark.
%
%
%
%
%

Now, we present the estimate of $m_0^2$ using the bare results
in SU(3)$_c$ lattice QCD as
%
%
%
%
\begin{eqnarray}
m_0^2 \equiv \qGq / \qq  \simeq  2.5\ {\rm GeV}^2 \qquad (\beta = 6.0 \ \ {\rm or} \ \ a^{-1} \simeq 2{\rm GeV}).
\end{eqnarray}
The large value of $m_0^2$ suggests the importance of
the   mixed  condensate   in  OPE.
Although we  do not include  renormalization effect, 
this  result
itself is determined very precisely~\cite{DOIS:qGq}.



\section{Discussions and Outlook}
\label{sec:summary}

For comparison with the standard value in the QCD sum rule,
we change the renormalization point from  $\mu\simeq\pi/a$ to $\mu\simeq 1$
GeV corresponding to the QCD sum rule.
Following Ref.~\citen{K&S},  we  first  take  the lattice results  of  the
condensates  as  the starting  point  of  the  flow, then  rescale  the
condensates perturbatively.
We adopt the anomalous dimensions at the one-loop level~\cite{Narison2},
and choose the parameters $\Lambda_{\rm QCD} = 200-300 \mbox{MeV}$ and
$N_f=0$ corresponding to quenched lattice QCD.
We obtain
$ m_0^2  \big|_{\mu=1{\rm GeV}} \equiv  \qGq/\qq  \big|_{\mu=1{\rm GeV}} \sim  3.5-3.7 \  
{\rm GeV}^2$.
%
%
%
%
%
%
Comparing with the standard value of  $m_0^2 = 0.8 \pm 0.2$ GeV$^2$ in the
QCD sum rule,  our calculation results in a  rather large value. (Note
that  the instanton  model has  made  a slightly  larger estimate  as
$m_0^2 \simeq 1.4$ GeV$^2$ at $\mu\simeq 0.6$ GeV~\cite{Polyakov}.)
For a more definite determination of $m_0^2$, the non-perturbative
renormalization scheme may be desired.

\begin{figure}[tbh]
\begin{center}
\includegraphics[scale=0.29]{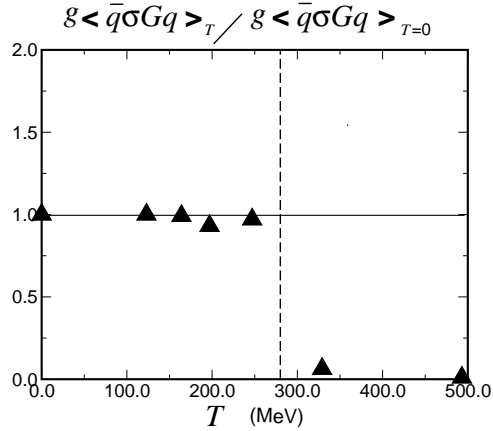}
\caption{\label{fig:qGq_finite_T}
The quark-gluon mixed condensate 
$\qGq_T/\qGq_{T=0}$
%
%
plotted against the temperature $T$.
The jackknife errors are hidden in the triangles.
The vertical dashed line denotes the critical temperature $T_c \simeq 280{\rm MeV}$ at the quenched level.
}
\end{center}
\end{figure}

We  again  emphasize  that  the  mixed condensate  $\qGq$  plays  the
important roles  in various contexts in quark  hadron physics.  
Hence, it is preferable to perform  further studies. 
In particular, the  thermal  effects  are interesting in relation 
to chiral restoration, because the  mixed  condensate is  another 
chiral  order parameter.  
Considering also the on-going experiments for the finite-temperature QCD in the RHIC project, 
we investigate the thermal effects on the mixed condensate $\qGq$.
We perform the calculation using the $16^3\times N_t$ lattices 
with $N_t=16, 12, 10, 8, 6, 4$ at $\beta=6.0$.
Figure~\ref{fig:qGq_finite_T} shows the lattice QCD results for the mixed 
condensate at finite temperature.
We find a drastic change of the mixed condensate around the critical temperature $T_c$, 
which reflects chiral-symmetry restoration~\cite{DOIS:T}.

In summary, 
we  have studied  the quark-gluon  mixed condensate  $\qGq$  using 
SU(3)$_c$ lattice QCD with  the KS fermion at the quenched
level.
For each  quark mass of  $m_q=21, 36, 52$  MeV, we have  generated 100
gauge configurations  on the $16^4$ lattice with  $\beta =6.0$.
Using the 1600 data for each $m_q$,  
we have found $m_0^2 \equiv \qGq / \qq
\simeq 2.5$ GeV$^2$ in the chiral limit at the lattice scale corresponding to 
$\beta=6.0$ or $a^{-1} \simeq$ 2 GeV.
We have also investigated $\qGq$ at finite temperature in lattice QCD.


\section*{Acknowledgements}

We would like to thank Dr. H. Matsufuru for his useful comments on 
the programming technique. 
This work is supported in part by the Grant for Scientific Research 
(No.11640261, No.12640274 and No.13011533) 
from the Ministry of Education, Culture, Science and Technology, Japan.
T.D. acknowledges the support of 
the JSPS Research Fellowships for Young Scientists.
The Monte Carlo simulations have been performed on the 
NEC SX-5 supercomputer at Osaka University.

%

\end{document}